 \documentclass[final,5p,times,twocolumn,authoryear]{elsarticle}

\usepackage{amsmath}
\usepackage{amssymb}
\usepackage{lipsum}
\usepackage{bm}
\usepackage{CJK}
\usepackage{url}
\usepackage{color}
\usepackage{booktabs}
\usepackage{bookmark}
\usepackage{natbib}
\usepackage{hyperref}
\usepackage{cleveref}
\usepackage{tikz}

\journal{Physics Letters B}

\begin{document}

\begin{frontmatter}



\title{Quantifying the Information Gain from Future High-Precision Radius Measurements for Identifying Twin Neutron Stars}


\author{Bao-An Li and Xavier Grundler}
\affiliation{organization={Department of Physics and Astronomy, East Texas A\&M University}, 
city={Commerce}, 
state={Texas},
postcode={75429-3011}, 
country={USA}}
\begin{abstract}
Twin neutron stars (NSs), characterized by identical gravitational masses but different radii, are among the most promising astrophysical signatures of a strong first-order hadron--quark phase transition in supradense matter. We investigate how increasingly precise NS radius measurements improve the Bayesian inference of twin-star observability using mock radius data for a canonical $1.4\,M_\odot$ NS. Radius uncertainties are varied from the current level of about $0.9$ km to the $\approx 0.1$ km precision anticipated from future X-ray and gravitational-wave observations. We quantify the information gained using the posterior distribution of the maximum twin-star radius separation $\Delta R$ together with an analytical model of branch distinguishability and complementary information-theoretic measures based on the branch observational efficiency and the Shannon entropy. The combined analyses reveal three inference regimes: a prior-dominated regime for $\sigma_R \gtrsim 0.6$ km, a rapid information-gain regime for $0.2 \lesssim \sigma_R \lesssim 0.6$ km, and an information-saturation regime for $\sigma_R \lesssim 0.2$ km. These complementary analyses consistently indicate that radius measurements with a precision of about $0.2$ km already extract most of the information available for identifying twin NSs within the present Bayesian framework. Beyond establishing a quantitative observational benchmark for future high-precision radius measurements, this work provides a general Bayesian framework for quantifying the information gain from progressively more precise observations and identifying the point of diminishing scientific returns.
\end{abstract}




\end{frontmatter}



\section{Introduction}\label{intro}

Twin neutron stars (NSs) are two compact stars with the same
gravitational mass but different radii.
Their possible existence was proposed decades ago
\cite{gerlach_1968,Kampfer:1981yr,Glendenning:1998ag,Schertler_2000}
as a natural consequence of a strong first-order phase transition in the
equation of state (EOS) of supradense matter.
Depending on the formation history, two stars may be born as twins or
evolve into twin configurations with identical masses but different
internal compositions and radii
\cite{Bejger:2016emu,Chanlaridis:2024rov}.
The existence, composition, and observational signatures of twin stars
are therefore intimately connected to the poorly known EOS of
supranuclear matter
\cite{Alford:2013aca,Blaschke:2013ana,Benic:2014jia,Kaltenborn:2017hus,Christian_2018,Alvarez-Castillo:2018pve,Haque26,Haque:2026dre}.

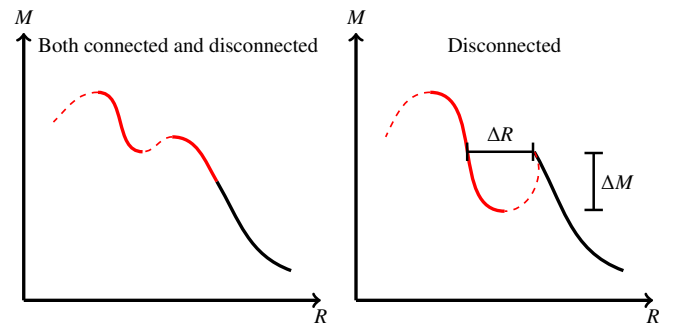
\begin{figure}
\resizebox{\linewidth}{!}{
\begin{tikzpicture}
    \node at (2.6,4.3) {Both connected and disconnected};

    \draw[ultra thick, ->] (0,0) -- (5,0) node[below] {$R$};
    \draw[ultra thick, ->] (0,0) -- (0,4.5) node[above] {$M$};

    \draw[ultra thick] (4.5,0.5) to[out=160,in=180+120] (3.25,2);
    \draw[ultra thick,red] (3.25,2) to[out=120,in=180+180] (2.5,2.75);
    \draw[thick,red,dashed] (2.5,2.75) to[out=180,in=180+180] (2,2.5);
    \draw[ultra thick,red] (2,2.5) to[out=180,in=180+180] (1.25,3.5);
    \draw[thick,red,dashed] (1.25,3.5) to[out=180,in=180+225] (0.5,3);
\end{tikzpicture}
\begin{tikzpicture}
    \node at (2.5,4.3) {Disconnected};

    \draw[ultra thick, ->] (0,0) -- (5,0) node[below] {$R$};
    \draw[ultra thick, ->] (0,0) -- (0,4.5) node[above] {$M$};

    \draw[ultra thick] (4.5,0.5) to[out=160,in=180+120] (3,2.5);
    \draw[thick,red,dashed] (3,2.5) to[out=180+120,in=180+180] (2.5,1.5);
    \draw[ultra thick,red] (2.5,1.5) to[out=180,in=180+180] (1.25,3.5);
    \draw[thick,red,dashed] (1.25,3.5) to[out=180,in=180+245] (0.5,2.75);

    \draw[very thick,|-|] (4,2.5) -- (4,1.5) node[midway,xshift=0.4cm] {$\Delta M$};
    \draw[very thick,|-|] (3,2.5) -- (1.85,2.5) node[midway,yshift=0.3cm] {$\Delta R$};
\end{tikzpicture}
}
\caption{Sketch modified from plots in Refs. \cite{Alford:2013aca,Christian_2018,ZhangLi_twin,Grundler:2025mcz,Haque:2026dre}. The change from black to red marks the appearance of QM in the core. Dashed lines represent unstable configurations.}
\label{fig:cat}
\vspace{-0.45cm}
\end{figure}

\begin{table*}
\vspace{-1.8cm}
 \caption{Representative distributions of mass-radius topologies predicted by posterior EOSs within a Bayesian framework \cite{Grundler:2025mcz} using mock radius data $R_{1.4}=11.9\pm \sigma_R$ (km) as well as currently available constraints on NS maximum mass, causality, and stability. Accepted EOSs in each category from 12 Markov Chain Monte Carlo (MCMC) walkers going for 300,000 steps ($\sigma = 0.9$ km), 16 walkers for 300,000 ($\sigma = 0.5$ km), and 36 walkers for 600,000 after 30,000 burn-in steps($\sigma = 0.1$ and 0.2 km).}
    \centering
    \begin{tabular}{|c|c|c|c|c|}
        \hline
         & $\sigma_R = 0.9$ & $\sigma_R = 0.5$ & $\sigma_R = 0.2$ & $\sigma_R = 0.1$ km\\
        \hline
        Absent & 119,179 & 92,710 & 135,085 & 81,510\\
        Both & 76,265 & 58,481 & 86,845 & 53,293\\
        Connected & 894,892 & 797,286 & 1,454,677 & 945,159\\
        Disconnected & 99,227 & 78,072 & 106,876 & 61,031\\
        All accepted & 1,189,563 & 1,026,549 & 1,783,483 & 1,140,993\\
        Twin NSs \% & 14.75 & 13.30 & 10.86 & 14.04\\
        \hline
    \end{tabular}\label{Truns}
\end{table*}

Despite extensive theoretical and observational efforts over the past
three decades, no convincing evidence for NS twins has yet
been established.
One major obstacle is the limited precision of current radius
measurements, which makes it difficult to distinguish the different
branches of the mass--radius ($M$--$R$) relation predicted by EOSs with
first-order hadron--quark phase transitions.
Fig.~\ref{fig:cat} illustrates two representative topologies of the
$M$--$R$ relation.
The observability of twin stars is commonly characterized by two
quantities:
the mass interval
$\Delta M$ separating the two stable branches and the maximum radius
difference
$\Delta R$ between twin configurations
\cite{ZhangLi_twin,Grundler:2025mcz}.
While $\Delta M$ determines the mass range over which twins may occur,
the statistical resolvability of the two branches is ultimately governed
by the competition between the intrinsic radius separation $\Delta R$
and the observational uncertainty $\sigma_R$.
Consequently, the ratio $\Delta R/\sigma_R$ provides a natural measure
of the observability of twin-star signatures.

Recently, incorporating main features of a first-order
hadron--quark phase transition similar to Refs.~\cite{Alford:2013aca,zdunik2013,Alvarez_Castillo_2016,
Tsa23,Gorda_2023,Carlomagno:2023nrc,Jimenez:2024hib,Albino:2024ymc,Li_2024,Chanlaridis:2024rov,Veselsky:2024bnf,Christian_2025,Pal_2025,Laskos_Patkos_2025,huang2025,chunhuang_2025twin} within a meta-model EOS for NSs \cite{zhang2018combined,xie2020bayesian,XieLi_phasetrans,Zhang:2021xdt,
Li:2024imk,Li:2025tku},
we investigated the possibility of forming NS twins using both the forward modeling approach \cite{ZhangLi_twin} and Bayesian inference
\cite{Grundler:2025mcz} constrained by current observations of NS radii, tidal deformabilities, and the maximum mass. These studies demonstrated that existing observations do not exclude the existence of twin NSs but leave substantial uncertainties in their
predicted properties.

\begin{figure*}
\vspace{-1.8cm}
    \centering
    \begin{tabular}{ccc}
        \includegraphics[width=0.45\linewidth]{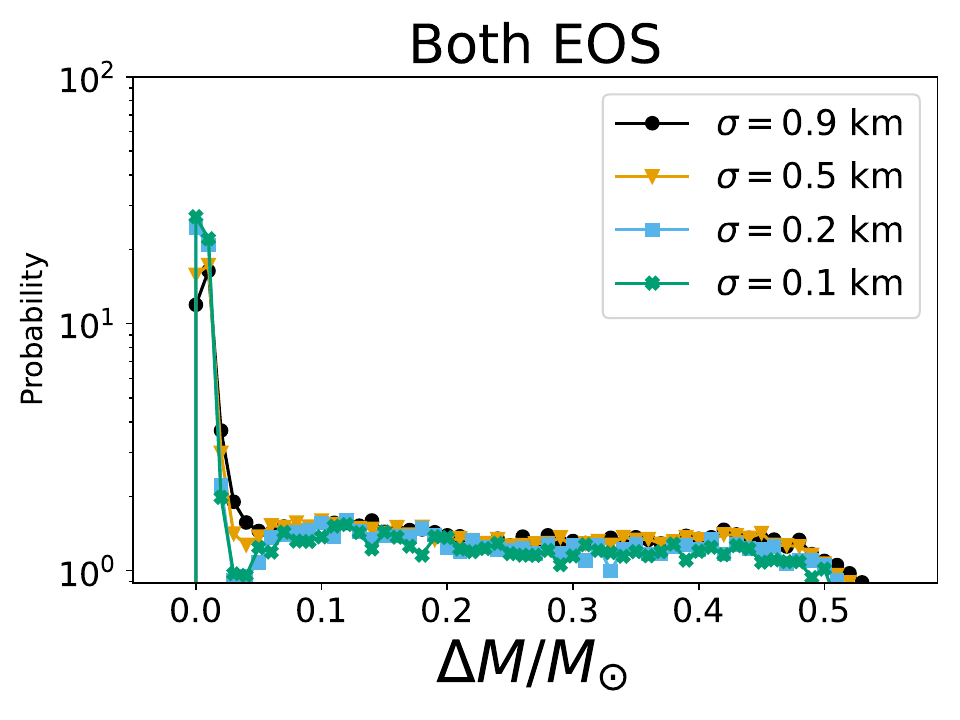} &
        \includegraphics[width=0.45\linewidth]{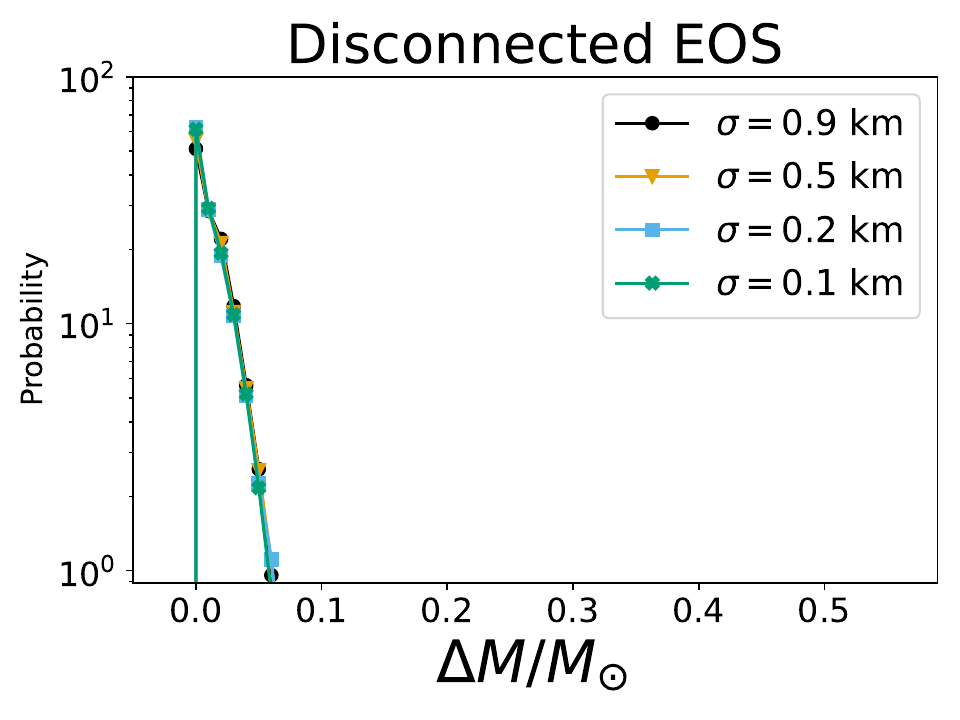}\\
        \includegraphics[width=0.45\linewidth]{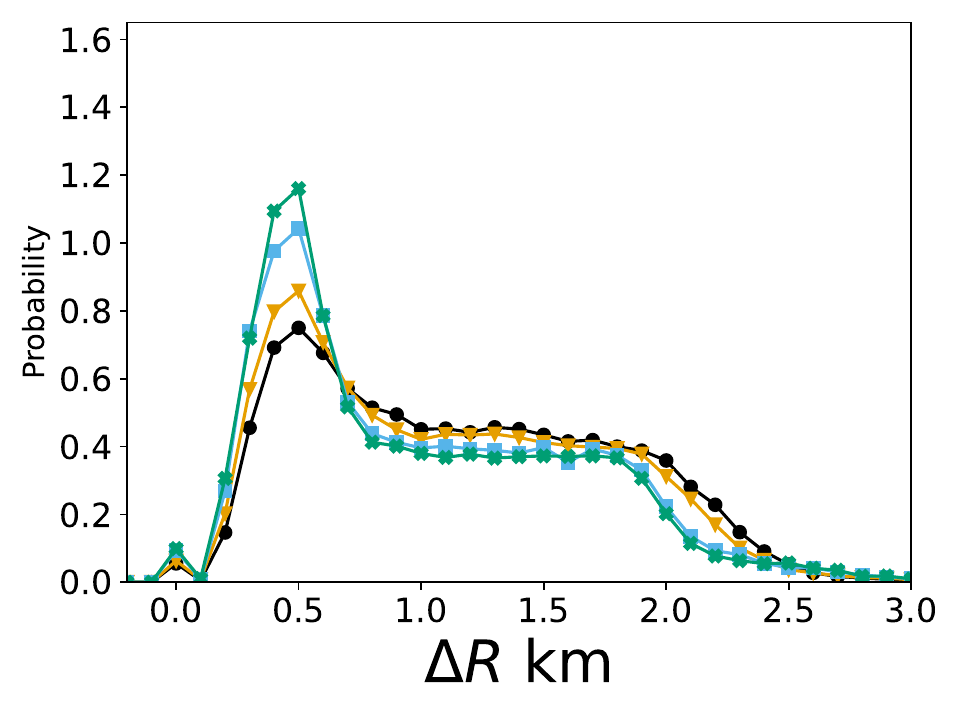} &
        \includegraphics[width=0.45\linewidth]{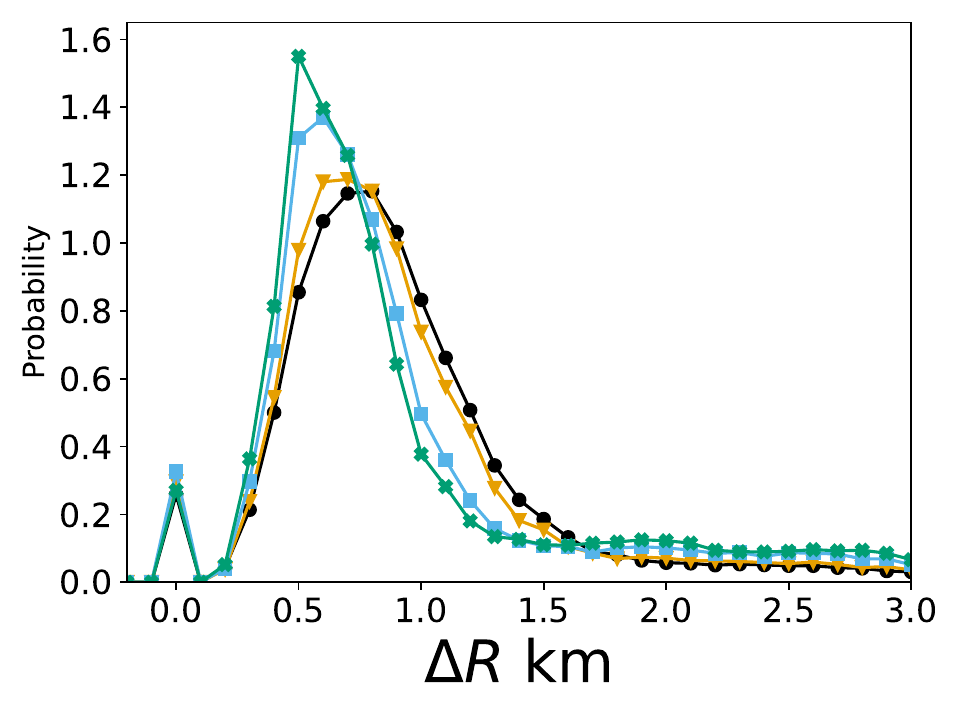}
    \end{tabular}
    \caption{Posterior probability distributions of $\Delta R$ and $\Delta M$ for Both (left) and Disconnected (right) categories 
from Bayesian inference using the mock radius data of $R_{1.4}=11.9\pm\sigma_R$ km data with $\sigma_R=0.1,0.2,0.5,0.9$ km, respectively.}\label{DrDm}
\end{figure*}

An important question is therefore:
\emph{What observational precision in NS radius measurements
is required to identify twin-star branches with high confidence?}
This question has become particularly timely because the next generation
of X-ray pulse-profile missions
\cite{eXTP,Ath},
together with third-generation gravitational-wave observatories,
including the Einstein Telescope
\cite{Sathyaprakash:2012jk},
Cosmic Explorer
\cite{Evans:2021gyd},
and other planned facilities
\cite{Hild:2009ns,LIGOScientific:2020zkf,
Chatziioannou:2021tdi,Pacilio:2021jmq,
Bandopadhyay:2024zrr,Finstad:2022oni,Walker:2024loo},
is expected to improve the precision of NS radius
measurements by nearly an order of magnitude over current capabilities.
Determining the observational precision beyond which further improvements yield diminishing scientific returns
is therefore important not only for understanding supradense matter but also for guiding
the design of future observing facilities.

In this Letter, we develop a Bayesian framework for quantifying
the information gained from progressively improving NS
radius measurements and apply it to assess the observational
prospects for identifying twin NSs. Using Bayesian analyses with mock radius
measurements of varying precision, we combine the posterior
distribution of the maximum twin-star radius separation with an
analytical model of branch distinguishability and complementary
information-theoretic measures based on the Shannon entropy
and the branch observational efficiency. This unified framework
enables us to quantify the evolution of observational information
and to determine the measurement precision beyond which
further improvements yield only diminishing scientific returns.
We show that the information extracted from radius
measurements approaches saturation once the observational
precision reaches approximately $\sigma_R \approx 0.2$ km,
thereby establishing a quantitative observational benchmark
for future high-precision studies of twin NSs.

\section{Bayesian framework and EOS parametrization}
\label{method}

This work is carried out within the Bayesian framework for identifying twin NSs developed in
Ref.~\cite{Grundler:2025mcz}, where the hadron--quark transition density
is assigned a uniform prior over $(1$--$6)\rho_0$, with
$\rho_0=0.16~\mathrm{fm}^{-3}$ denoting the nuclear saturation density.
Rather than using only the currently available radius constraints from
LIGO/Virgo and NICER, we employ mock radius measurements for a canonical
NS, $R_{1.4}=11.9~\mathrm{km}\pm\sigma_R$,
with $\sigma_R$ varying from $0.9$ km, corresponding approximately to
the present uncertainty inferred from GW170817 \cite{2018gw170817}, down to $0.1$ km, representative of the
precision anticipated from future observations. This approach enables a
systematic investigation of the information gained as the observational
precision improves.

We briefly summarize here the EOS parametrization and Bayesian framework
used to generate the ensemble of NS EOSs considered in this work; detailed descriptions can be found in
Refs.~\cite{zhang2018combined,Zhang:2019fog,Zhang:2021xdt,
xie2019bayesian,xie2020bayesian,xie2021bayesian,Zhang:2023wqj,
Zhang:2024npg,Xie:2024mxu,Li:2024imk,LiEPJA,universeReview,
XieLi_phasetrans,Li:2025tku}.  The EOS of charge neutral hadronic matter (HM)
consisting of neutrons, protons, electrons, and muons in
$\beta$ equilibrium is constructed using the empirical
isospin-parabolic form \cite{bombaci1991asymmetric}
\begin{equation}
E(\rho,\delta)=E_0(\rho)+E_{\rm sym}(\rho)\delta^2+
{\cal O}(\delta^4),
\end{equation}
for the energy per nucleon in neutron-rich matter at density $\rho=\rho_n+\rho_p$ and isospin asymmetry 
$\delta=(\rho_n-\rho_p)/\rho$.  The symmetric nuclear matter (SNM) energy $E_0(\rho)$ and
symmetry energy $E_{\rm sym}(\rho)$ are parameterized as 
\begin{align}
E_0(\rho) &=
E_0(\rho_0)
+\frac{K_0}{2}x^2+\frac{J_0}{6}x^3,\\
E_{\rm sym}(\rho) &=
E_{\rm sym}(\rho_0)
+Lx+\frac{K_{\rm sym}}{2}x^2
+\frac{J_{\rm sym}}{6}x^3,
\end{align}
where
$x=(\rho-\rho_0)/(3\rho_0)$ and $E_0(\rho_0)=-16$~MeV.
The five coefficients
$K_0$, $J_0$, $L$, $K_{\rm sym}$, and $J_{\rm sym}$, together with
$E_{\rm sym}(\rho_0)$, characterize the hadronic EOS. They are generated randomly within their prior ranges uniformly.  
These polynomial forms serve as flexible parameterizations rather than
Taylor expansions of any particular energy-density functional known \cite{zhang2018combined} and
therefore are used as meta-models over the supra-saturation density
range explored in the Bayesian analysis \cite{xie2019bayesian,xie2020bayesian,Grundler:2025mcz}.

Possible deconfinement is introduced through a Maxwell construction
to a constant-speed-of-sound (CSS) quark-matter (QM) EOS ~\cite{Alford:2013aca,zdunik2013}, where the energy density $\varepsilon(p)$ is parameterized as a function of pressure $p$
\begin{equation}
\varepsilon(p)=
\begin{cases}
\varepsilon_{\rm HM}(p), & p<p_t,\\
\varepsilon_t+\Delta\varepsilon+
c_{\rm qm}^{-2}(p-p_t), & p>p_t.
\end{cases}
\end{equation}
The three additional parameters are the transition density
$\rho_t/\rho_0$, the relative energy-density discontinuity
$\Delta\varepsilon/\varepsilon_t$, and the squared QM sound speed
$c_{\rm qm}^2/c^2$.  Thus, the full EOS ensemble is controlled by
nine empirical parameters,
\begin{equation}
\boldsymbol{\theta}=
(E_{\rm sym}(\rho_0),K_0,J_0,L,K_{\rm sym},J_{\rm sym},
\rho_t/\rho_0,\Delta\varepsilon/\varepsilon_t,c_{\rm qm}^2/c^2).
\end{equation}
The CSS parameterization provides a simple representation of the
high-density QM sector while allowing the transition density, its
strength, and the stiffness of QM to vary independently.

We sample the nine parameters over their adopted prior ranges consistent with the latest terrestrial laboratory experiments \cite{Grundler:2025mcz} and then solve the TOV equations \cite{tolman1939,oppenheimer1939massive} for each EOS. 
For each sampled EOS, as part of the likelihood function in our Bayesian analyses, we impose the physical conditions of a positive
crust--core transition pressure, thermodynamic stability, and causality,
as well as the conservative maximum-mass constraint
$M_{\rm TOV}\geq1.97\,M_\odot$.  The remaining likelihood $P_R(\rm{theory},\rm{given~ the~ data})$ is based on
a single mock canonical-radius measurement,
$R_{1.4}=11.9\pm\sigma_R$ km, with
\begin{equation}
P_R \propto
\exp\left[-\frac{(R_{1.4}^{\rm th}-11.9\,{\rm km})^2}
{2\sigma_R^2}\right],
\end{equation}
where, for EOSs supporting twin configurations, the branch whose
predicted radius is closest to the mock measurement is used.  We vary
$\sigma_R$ to quantify how improving the precision of a single
canonical-radius measurement changes the posterior distributions of
the EOS and twin-star properties. Consequently, all posterior
constraints discussed below are conditional on the adopted
meta-model EOS, prior ranges, physical and stellar constraints, and the
NS radius likelihood $P_R$.  The quantitative precision scale found in this
work should therefore be understood within this Bayesian EOS
framework rather than as a model-independent threshold.

\section{Results and discussion}\label{results}

\begin{figure}[thb]
\centering
 \resizebox{0.5\textwidth}{!}{
\includegraphics[width=0.7\textwidth]{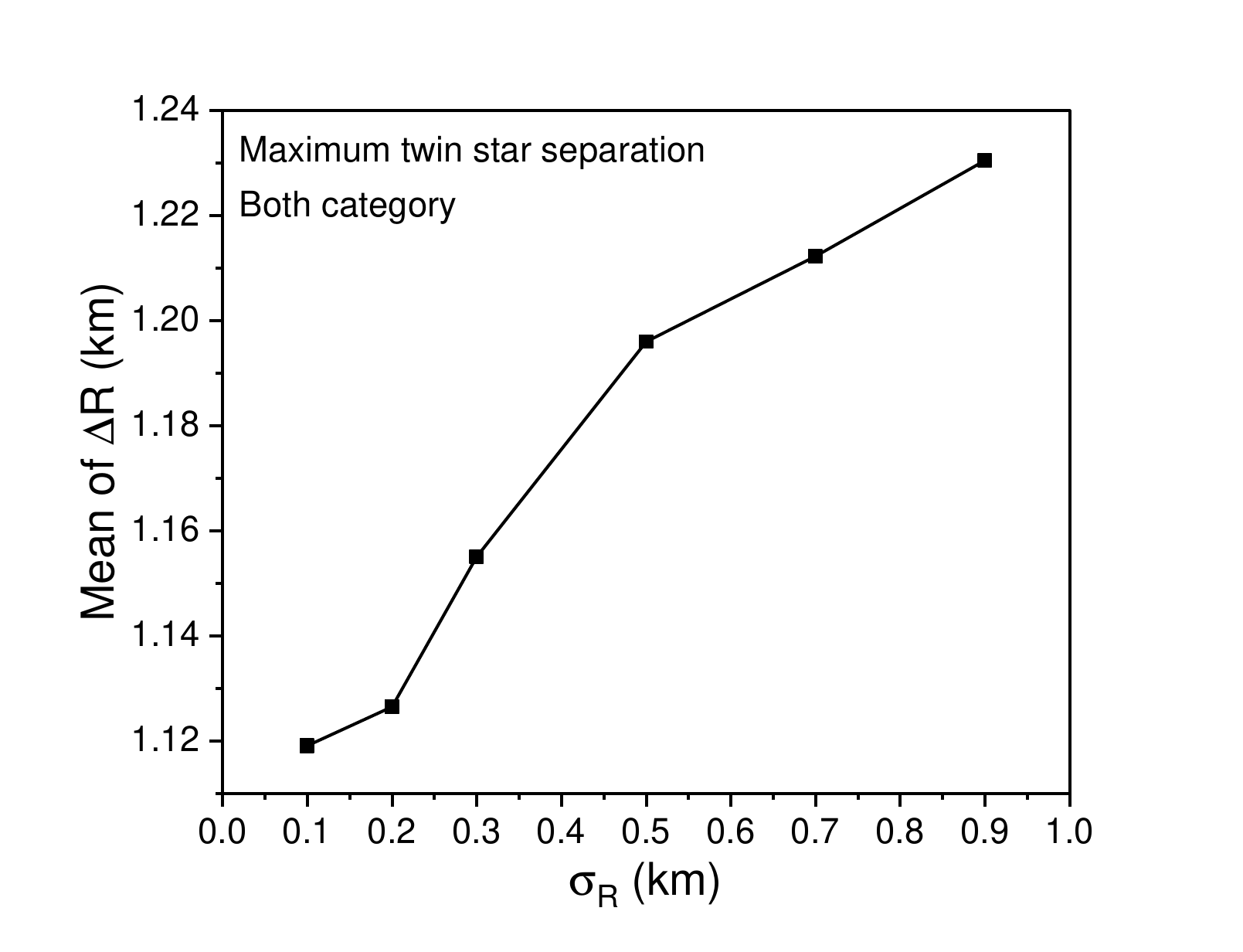}
}
\caption{Posterior mean of the maximum radius separation $\Delta R$ of twin NSs as a function of the assumed
radius uncertainty $\sigma_R$ for the Both category. } \label{mean}
\end{figure}

The posterior mass--radius sequences predicted by the accepted EOSs are
classified according to the topology scheme proposed in
Ref.~\cite{Alford:2013aca}: Absent, Connected, Both, or
Disconnected. As illustrated in
Fig.~\ref{fig:cat}, the two categories most relevant for investigating twin NSs are: (i) the \emph{Disconnected} category, in which the phase
transition immediately destabilizes the star on the first branch before
stability is recovered on a separate second branch, and (ii) the
\emph{Both} category, in which the phase transition initially preserves
stability, followed by an unstable region and a subsequent disconnected
branch. Representative posterior statistics are summarized in
Table~\ref{Truns}. Depending on the assumed observational precision,
EOSs producing twin NSs (both and disconnected categories sketched in Fig. \ref{fig:cat}) account for approximately
$10\%$--$15\%$ of all accepted EOSs. In contrast, the \emph{Absent}
category contains EOSs for which either the hadron--quark phase
transition does not occur within stable NSs or the onset of
quark matter immediately destabilizes the star without re-establishing a
second stable branch. Consequently, no stable NS in this
category contains a quark-matter core.

\subsection{Precision-dependence of the observability of twin stars}\label{PreD}

Fig.~\ref{DrDm} presents the posterior probability distributions
$P(\Delta R,\sigma_R)$ and $P(\Delta M,\sigma_R)$ for the Both (left)
and Disconnected (right) categories obtained from Bayesian analyses
using the mock radius constraint
$R_{1.4}=11.9\pm\sigma_R$ km with
$\sigma_R=0.1$, $0.2$, $0.5$, and $0.9$ km.
Several noteworthy features emerge.

First, the posterior distributions of the maximum radius separation
$\Delta R$ become progressively narrower as the radius uncertainty
decreases, indicating that improved observations provide increasingly
stronger constraints on the intrinsic separation between the two
branches. Most of the posterior narrowing occurs when the observational
precision increases from $\sigma_R=0.9$ km to about $0.2$ km, whereas
further reductions to $0.1$ km produce only modest changes. Fundamentally, this is because the posterior distribution functions of the underlying EOS parameters have essentially all reached saturation once $\sigma_R$ becomes smaller than about 0.2 km as demonstrated in our earlier work \cite{Li:2024imk,Grundler:2025mcz,Li:2025tku}.

Second, the Disconnected category, corresponding to twins consisting of
one purely hadronic star (made of neutrons, protons, electrons and muons \cite{Grundler:2025mcz}) and one hybrid star, exhibits a sharply peaked distribution centered at
$\Delta R\approx0.5$ km.
In contrast, the corresponding posterior distribution of the maximum
mass gap $\Delta M$ remains nearly unchanged with observational
precision and is strongly concentrated around $\Delta M\approx0$,
indicating that the radius separation is considerably more sensitive
than the mass gap to improvements in observational accuracy of NS radius.

Third, for the Both category, which contains both hadronic-hybrid and hybrid--hybrid twin stars,
both $P(\Delta R,\sigma_R)$ and $P(\Delta M,\sigma_R)$ display broad
distributions with long shoulders or tails extending to larger values, reflecting the greater diversity of
EOSs capable of producing connected and disconnected hybrid-star
branches. The long tails of these distributions exhibit only weak dependence on $\sigma_R$.

An important distinction emerges between the location of the dominant
$\Delta R$ population and its posterior mean  
\begin{equation}
\langle \Delta R\rangle(\sigma_R)
=
\int \Delta R\cdot P(\Delta R,\sigma_R)\cdot d\Delta R.
\end{equation}
For the Both category, the most probable value of $\Delta R$ remains close to
$\sim 0.5$~km as $\sigma_R$ is varied, whereas the posterior mean
changes more appreciably.  This difference arises from the broad
high-$\Delta R$ shoulder of the distribution: although this shoulder
contains relatively little probability compared with the dominant
peak, its contribution is amplified in the posterior mean by the
factor of $\Delta R$ in the averaging.  Improving the radius precision
therefore primarily redistributes the posterior weight among the
large-$\Delta R$ configurations rather than substantially shifting the
most probable twin-star separation.

The Disconnected category exhibits a qualitatively different behavior.
Its $\Delta R$ distribution is sharply concentrated around its dominant
peak and contains only a weak high-$\Delta R$ tail.  Consequently, its
posterior mean is close to the location of the dominant peak and changes
only weakly with $\sigma_R$, even though the peak itself exhibits a
small displacement.  The weak precision dependence of both the mean
and the entropy in this category indicates that the intrinsic radius
separation of the accepted Disconnected EOSs is already relatively
well localized.  Thus, the two topologies illustrate an important
distinction between constraining the intrinsic twin-star separation and
improving the observational ability to distinguish the two branches.

The Both category also exhibits an extended high-$\Delta M$ tail,
whereas the corresponding distribution for the Disconnected category
is sharply concentrated near $\Delta M\simeq 0$.  The high-$\Delta M$
tail indicates that a subset of the accepted EOSs in the Both category
supports a relatively broad mass interval over which the two stable
branches can coexist as twin configurations.  This reflects the larger
structural diversity of EOSs capable of producing both connected and
disconnected hybrid-star branches.  The present analysis does not,
however, isolate a unique microscopic mechanism or EOS parameter responsible for
this tail; establishing such a correlation would require a dedicated
mapping of $\Delta M$ onto the hadron-quark transition and high-density EOS
parameters.

Fig.\ref{mean} further illustrates the dependence of the inferred
maximum radius separation on the observational precision by showing the
posterior mean of $\Delta R$ as a function of $\sigma_R$ for the Both category. It reveals three distinct regimes in the dependence of the posterior mean of $\Delta R$ on the assumed
radius uncertainty $\sigma_R$. For large uncertainties
($\sigma_R\gtrsim0.6$ km), the posterior mean changes only weakly with
$\sigma_R$, indicating that the Bayesian inference is {\it largely
prior-dominated} because the observational constraint is too weak to
significantly discriminate among competing EOSs. As the radius precision
improves into the intermediate range
($0.2\lesssim\sigma_R\lesssim0.6$ km), the posterior responds most rapidly to
the increasing constraining power of the NS radius data, leading to the {\it greatest
information gain}. Once the observational precision reaches
approximately $\sigma_R\approx0.2$ km, the curve flattens again,
signaling an {\it information-saturation regime} in which further reductions
in $\sigma_R$ produce only marginal changes in the inferred maximum
radius separation $\Delta R$. 

Thus, Fig.~\ref{mean} illustrates the transition from
a prior-dominated regime to a likelihood-dominated regime, with the
crossover occurring around $\sigma_R\sim0.2$--$0.6$ km. It suggests that radius measurements with
a precision of about $0.2$ km are sufficient to extract most of the
information on twin-star observability contained within the present
Bayesian framework.

\subsection{Analytical interpretation of branch distinguishability}

\begin{figure}[thb]
\centering
\resizebox{0.5\textwidth}{!}{
\includegraphics[width=0.7\textwidth]{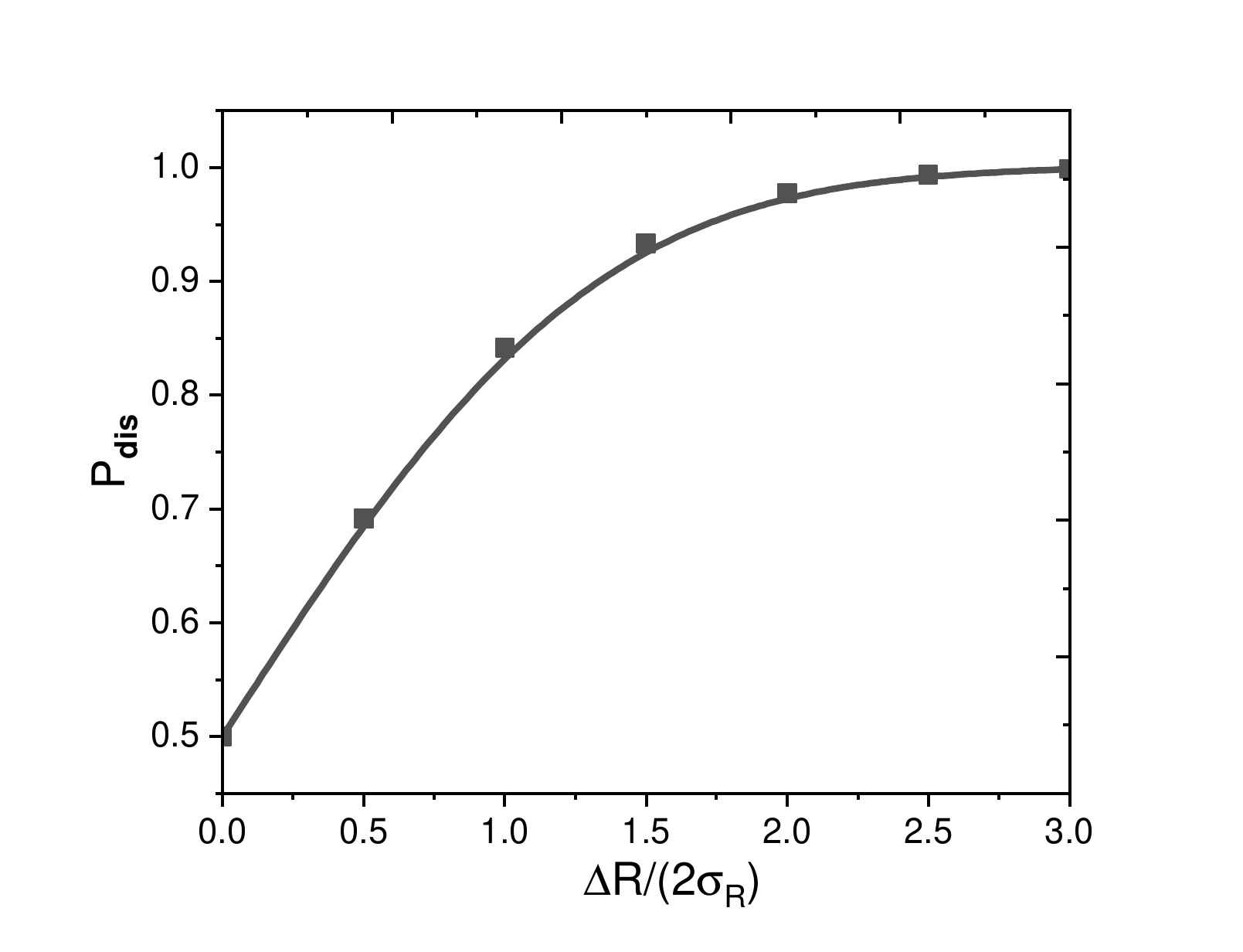}
}
\caption{The branch distinguishability probability $P_{\rm dis}$ as a
function of the ratio $\Delta R/(2\sigma_R)$.}
\label{pdis}
\end{figure}

For two independent radius measurements with identical Gaussian
uncertainties $\sigma_R$, the uncertainty of their radius difference is
$\sigma_{\Delta R}=\sqrt{2}\sigma_R$.
Although this immediately suggests that larger values of
$\Delta R/\sigma_R$ improve the separation of two stellar
configurations, the identification of twin NSs relies on
Bayesian inference of the underlying mass--radius relation from an
ensemble of observations rather than on direct comparisons of individual
stellar pairs. It is therefore useful to introduce an analytical measure
of the statistical distinguishability of the two branches and compare it
with the Bayesian results.

Assuming Gaussian observational uncertainties, we model the inferred radii of two stellar configurations as Gaussian
random variables
$
R_1\sim\mathcal{N}(\mu_1,\sigma_R),\;
R_2\sim\mathcal{N}(\mu_2,\sigma_R),
$
with intrinsic radius separation
$\Delta R=|\mu_2-\mu_1|$.
For equal prior probabilities, the Bayes-optimal decision boundary is
located at the midpoint
$R_*=(\mu_1+\mu_2)/2$,
leading to the probability of correctly assigning an observed radius to
its corresponding branch \cite{Elements,Pattern},
\begin{equation}
P_{\rm dis}
=
\Phi\!\left(\frac{\Delta R}{2\sigma_R}\right),
\label{eq:Pdis}
\end{equation}
where
\begin{equation}
\Phi(z)
=
\frac{1}{\sqrt{2\pi}}
\int_{-\infty}^{z}
e^{-t^2/2}\,dt
\end{equation}
is the cumulative distribution function of the standard normal
distribution.

An equivalent interpretation is obtained from the overlap coefficient
\begin{equation}
\mathcal{O}
=
\int_{-\infty}^{\infty}
\min[p_1(R),p_2(R)]\,dR,
\end{equation}
which measures the common probability mass shared by the two
distributions. For Gaussian distributions with equal variance,
\begin{equation}
\mathcal{O}
=
2\,\Phi\!\left(-\frac{\Delta R}{2\sigma_R}\right),
\end{equation}
so that
\begin{equation}
P_{\rm dis}=1-\frac{\mathcal{O}}{2}.
\end{equation}
The overlap coefficient therefore provides an intuitive interpretation
of Eq.~(\ref{eq:Pdis}): as the overlap between the two radius
distributions decreases with increasing $\Delta R/\sigma_R$, the
probability of correctly identifying the corresponding branch
approaches unity.

As shown in Fig.~\ref{pdis}, the branch distinguishability depends only
on the dimensionless ratio $\Delta R/(2\sigma_R)$.
When $\Delta R/(2\sigma_R)\ll1$, the two radius distributions strongly
overlap and are difficult to distinguish statistically.
For $\Delta R/(2\sigma_R)\approx1$, the branches become moderately
resolvable, whereas for $\Delta R/(2\sigma_R)\gtrsim2$, the overlap
becomes negligible and $P_{\rm dis}$ approaches unity, indicating nearly
complete statistical resolution.

For representative twin NSs with $\Delta R\approx0.5$ km and
$\sigma_R\approx0.2$ km, Eq.~(\ref{eq:Pdis}) gives
$P_{\rm dis}=\Phi(1.25)\approx0.89$, which already lies close
to the saturation region shown in Fig.~\ref{pdis}. Thus, once
the observational precision reaches $\sigma_R\approx0.2$ km,
the two branches can already be identified with high
probability, and further improvements in radius precision yield
only marginal reductions in the uncertainty of the inferred
maximum radius separation $\Delta R$. This
information-saturation threshold is fully consistent with the
Bayesian results presented in the previous subsection: once
$\sigma_R\lesssim0.2$ km, not only the posterior distribution
of $\Delta R$ but also those of all nine EOS parameters exhibit
only weak additional evolution as the observational precision is
further improved \cite{Li:2024imk,Grundler:2025mcz,Li:2025tku}.

It is important to emphasize that the distinguishability considered here
does \emph{not} refer to resolving two individual NSs on the
sky. Twin stars may reside in completely different regions of the Galaxy
or even in different galaxies, making them trivially distinguishable by
their sky positions. Instead, we consider the \emph{statistical
distinguishability of the two branches in the mass--radius diagram}. The
quantity $P_{\rm dis}(\Delta R,\sigma_R)$ measures the probability that
the intrinsic radius separation exceeds the observational uncertainty
sufficiently for the two branches to be statistically resolved. It is
therefore a measure of the observational capability to identify two
distinct mass-radius branches rather than two spatially separated NSs.

The radius uncertainty $\sigma_R$ plays two distinct roles in the
present analysis. It first enters the Bayesian likelihood function,
where it determines the posterior distribution
$P(\Delta R,\sigma_R)$ by selecting EOSs predicting NS radii compatible with the mock radius
measurement. The same $\sigma_R$ subsequently appears in
Eq.~(\ref{eq:Pdis}) only as the observational uncertainty governing the
statistical resolvability of two branches for a given intrinsic radius
separation. These correspond to two different stages of the inference
process---the Bayesian determination of the posterior $\Delta R$ and the subsequent
assessment of mass-radius branch distinguishability.

\subsection{Information gain from high-precision NS radius data}
To quantify the information gained from increasingly precise radius
measurements in the search for twin NSs, we use two
complementary measures that characterize different aspects of the
Bayesian inference.

First, we define the two-branch observational efficiency as
\begin{equation}
\eta(\sigma_R)=
\int
P(\Delta R,\sigma_R)
P_{\rm dis}(\Delta R,\sigma_R)
\,d(\Delta R),
\label{eq:eta}
\end{equation}
which incorporates the entire inference process. Here
$P(\Delta R,\sigma_R)$ is the posterior probability distribution of the
intrinsic radius separation inferred from the Bayesian analysis, while
$P_{\rm dis}$ measures the probability that two branches with a given
separation can be statistically resolved. The efficiency
$\eta(\sigma_R)$ therefore represents the expected likelihood of
distinguishing the two branches for a specified observational
precision $\sigma_R$.

The second measure is the Shannon entropy \cite{Shannon1949,MacKay2003},
\begin{equation}
S(\sigma_R)
=
-\int
P(\Delta R,\sigma_R)
\ln[P(\Delta R,\sigma_R)]
\,d(\Delta R),
\label{eq:Shannon}
\end{equation}
which quantifies the residual uncertainty of the posterior distribution.
A smaller Shannon entropy corresponds to a more concentrated posterior distribution and therefore to greater information gained from the radius measurement.

\begin{figure}[thb]
\centering
\resizebox{0.5\textwidth}{!}{
\includegraphics[width=0.7\textwidth]{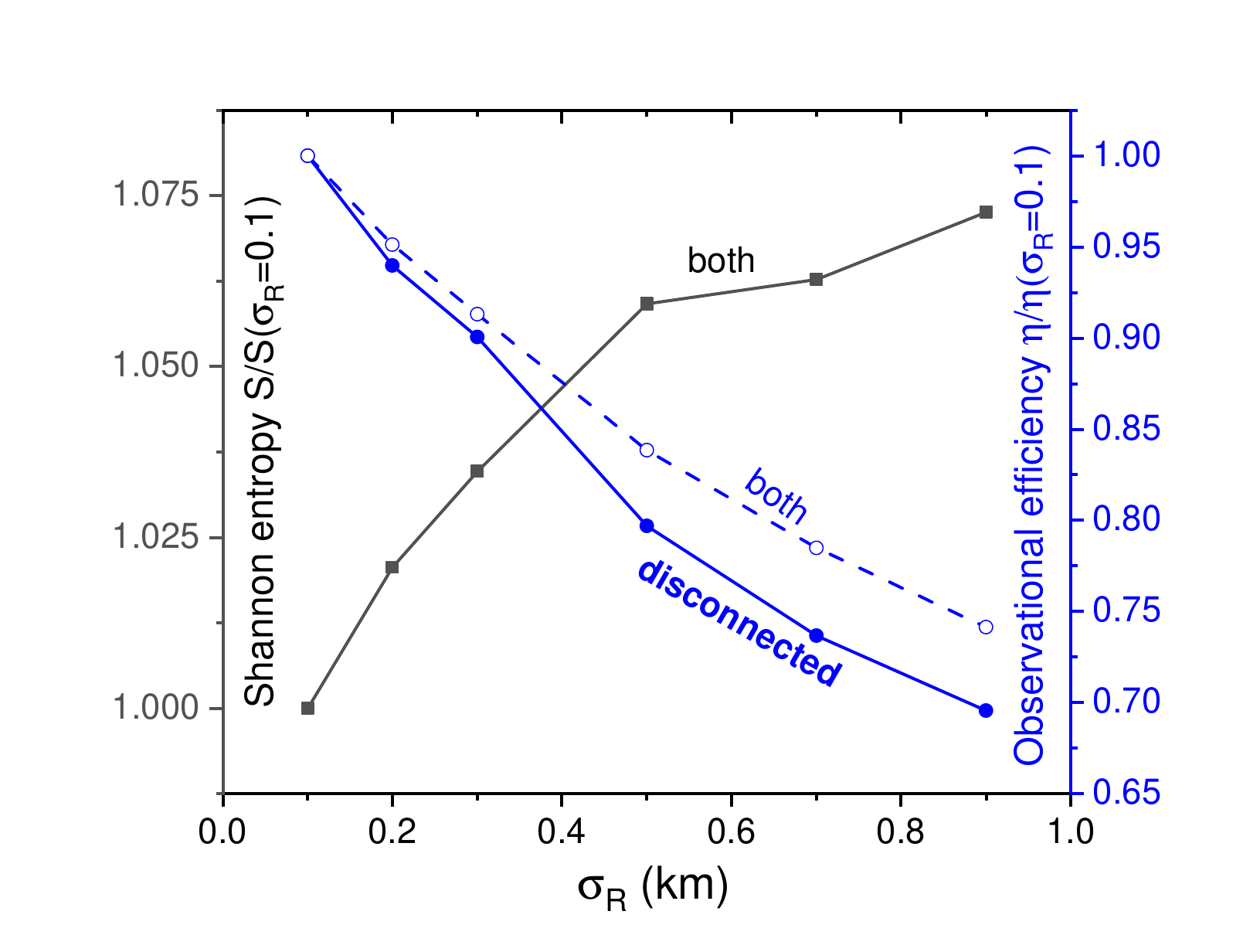}
}
\caption{Normalized Shannon entropy,
$S/S(\sigma_R=0.1)$ (black, left axis) for the Both category, and normalized branch observational efficiency,
$\eta/\eta(\sigma_R=0.1)$ (blue, right axis)  for both the disconnected and Both categories, as functions of the assumed
radius uncertainty $\sigma_R$.}
\label{entropy}
\end{figure}

Fig.~\ref{entropy} compares the normalized Shannon entropy with the normalized branch observational efficiency (with respect to the results using $\sigma_R=0.1$ km).
The two quantities characterize complementary aspects of the Bayesian inference.
The Shannon entropy measures the residual uncertainty of the posterior
distribution $P(\Delta R,\sigma_R)$, whereas the observational
efficiency measures the expected likelihood of statistically resolving the two mass-radius branches despite finite radius measurement uncertainties. 
The normalized Shannon entropy of the
$\Delta R$ distribution increases with increasing $\sigma_R$ for the Both
category, indicating a loss of information about the intrinsic
branch separation as the radius precision deteriorates. 

It is also seen from Fig.~\ref{entropy} that the observational efficiency for distinguishing the two branches decreases substantially with increasing
$\sigma_R$ for both the Disconnected and Both topologies.  Relative to its value at
$\sigma_R=0.1$ km, the efficiency decreases to about $0.74$ for the Both category and $0.70$ for the Disconnected category at
$\sigma_R=0.9$ km.  Thus, the weak precision dependence of the
intrinsic $\Delta R$ distribution in the Disconnected category (discussed in Section \ref{PreD}) does not imply weak precision dependence of its observational
distinguishability.  The latter is determined not only by the posterior
distribution of $\Delta R$ but also by the ability of a measurement
with finite radius uncertainty to resolve the two branches.  This
distinction is important: improved precision of NS radius measurements can substantially
enhance the detectability of twin branches even when it produces only
a modest change in the inferred intrinsic distribution of their radius
separation.

We emphasize that the quantities shown in Figs.~\ref{mean} and \ref{entropy} represent different functionals of
the same posterior distribution. Fig. \ref{mean} shows the first moment
(posterior mean) of $\Delta R$, whereas the Shannon entropy and the
branch observational efficiency shown in Fig.~\ref{entropy} are nonlinear and
weighted functionals of the entire posterior distribution $P(\Delta R,\sigma_R)$. Consequently,
they emphasize different aspects of the posterior and need not display
identical features in their dependences on the observational uncertainty $\sigma_R$.

As the assumed radius uncertainty increases, the normalized Shannon
entropy in the Both category increases monotonically, while the normalized branch
observational efficiency decreases monotonically, reflecting the gradual
loss of information contained in the posterior distribution. Although
their variations are smoother than those of the posterior mean in the Both category and the
analytical distinguishability probability, both quantities exhibit
distinct changes in slope with increasing $\sigma_R$. The entropy
increases most rapidly between $\sigma_R=0.1$ and $0.2$ km, followed by
a more gradual variation over $0.2\lesssim\sigma_R\lesssim0.6$ km,
before becoming only weakly dependent on $\sigma_R$ for
$\sigma_R\gtrsim0.6$ km. The observational efficiency exhibits a
corresponding evolution with progressively decreasing slope. These
complementary behaviors are consistent with the transition from the
information-rich regime to the prior-dominated regime established in
Figs.~\ref{mean} and~\ref{pdis}, while reflecting the fact that entropy and efficiency
probe the evolution of the entire posterior distribution rather than its
mean value.

We have also evaluated the posterior mean and Shannon entropy of
$\Delta R$ for the Disconnected category.  Both quantities exhibit
variations of less than about $3\%$ over the range
$0.1\leq\sigma_R\leq0.9$ km.  These changes are comparable to the
statistical fluctuations associated with the finite and
$\sigma_R$-dependent numbers of accepted MCMC samples (Table~\ref{Truns}).
Since the number of accepted EOSs cannot be fixed a priori to be the
same for different radius uncertainties, we do not regard these
few-percent variations as statistically significant and therefore
do not include them as separate curves in Figs.~\ref{mean} and \ref{entropy}. 

\section{Summary and conclusion}
Using Bayesian inference with mock NS radius measurements of progressively improving precision, we have quantified how future observations enhance the capability to identify twin NSs and determined the observational precision beyond which further improvements yield diminishing scientific returns.

Bayesian inference constrains the posterior distribution of the maximum radius separation $\Delta R$ progressively more tightly as the radius uncertainty decreases from the current level to about $0.2$ km, whereas the corresponding maximum mass separation $\Delta M$ remains comparatively insensitive to radius precision. The posterior mean of $\Delta R$ reveals three inference regimes: a prior-dominated regime for $\sigma_R \gtrsim 0.6$ km, the most rapid information-gain regime for $0.2 \lesssim \sigma_R \lesssim 0.6$ km, and an information-saturation regime for $\sigma_R \lesssim 0.2$ km. These results indicate that most of the information relevant to identifying twin NSs is acquired before the radius uncertainty reaches approximately $0.2$ km. Although the quantitative saturation threshold may depend somewhat on the adopted EOS parameterization and Bayesian framework, the existence of an information-saturation regime is expected to be a generic consequence of Bayesian inference with increasingly precise NS radius measurements.

An analytical model for branch distinguishability provides a physical explanation for this behavior. The distinguishability probability depends only on the ratio $\Delta R/(2\sigma_R)$ and approaches unity once the intrinsic branch separation significantly exceeds the observational uncertainty. For representative twin NSs with $\Delta R \approx 0.5$ km and $\sigma_R \approx 0.2$ km, the predicted distinguishability probability is already about 89\%, consistent with the onset of information saturation found in our Bayesian analysis.

Finally, the branch observational efficiency and the Shannon entropy provide complementary information-theoretic measures of the Bayesian inference. Their opposite but consistent trends show that increasing branch distinguishability is accompanied by decreasing posterior uncertainty, while also illustrating how different statistical functionals of the same posterior distribution respond differently to improvements in observational precision. Together, they provide independent support for the transition from the prior-dominated regime to the information-saturation regime.

Within the adopted Bayesian framework and EOS meta-model, our study indicates that NS radius measurements with a precision of approximately $0.2$ km are sufficient to extract most of the information available for identifying twin NSs. More broadly, the methodology developed here provides a general
framework for quantifying the scientific return of progressively
more precise observations and establishes a quantitative observational benchmark for future observational programs
designed to identify twin NSs and probe first-order hadron--quark phase transitions in supradense matter.
\\

\noindent{\bf Acknowledgement:} This work was supported in part by the U.S. Department of Energy, Office of Science, under Award Number DE-SC0013702, and the NASA-Texas Space Grant Consortium.

\bibliographystyle{elsarticle-harv} 
\bibliography{references}






\end{document}